\documentclass[aps,prl,groupedaddress,fleqn,twocolumn]{revtex4-1}
\pdfoutput=1
\usepackage{graphicx}
\usepackage{amsmath}
\usepackage{amssymb}
\usepackage{gensymb}

\begin{document}

\title{Frequency-locked chaotic opto-RF oscillator}

\author{Aur\'elien Thorette$^*$}
\author{Marco Romanelli}
\author{Marc Brunel}
\author{Marc Vallet}

\address{\small Institut de Physique de Rennes, Universit\'e Rennes I - CNRS UMR 6251 \\
	Campus de Beaulieu, 35042 Rennes Cedex, France	}
\address{$^*$Corresponding author: aurelien.thorette@univ-rennes1.fr}

\begin{abstract}
A driven opto-RF oscillator, consisting of a dual-frequency laser (DFL) submitted to frequency-shifted feedback, is studied experimentally and numerically in a chaotic regime. Precise control of the reinjection strength and detuning permits to isolate a parameter region of bounded-phase chaos, where the opto-RF oscillator is frequency-locked to the master oscillator, in spite of chaotic phase and intensity oscillations. Robust experimental evidence of this synchronization regime is found and phase noise spectra allows to compare phase-locking and bounded-phase chaos regimes. In particular, it is found that the long-term phase stability of the master oscillator is well transferred to the opto-RF oscillator even in the chaotic regime. 
\end{abstract}

\maketitle
	
Dual-frequency lasers (DFL) are convenient systems for the generation of continously tunable, high spectral purity, optically carried radiofrequency (RF) signals. Driven by applications such as microwave-photonic links, from radio-over-fibre to antenna feeds ~\cite{yao2009,kervella2014}, or highly monochromatic THz generation~\cite{rolland2014}, many solid-state DFLs have been investigated recently. This includes for instance diode-pumped solid-state lasers~\cite{pillet2014,miaohu2015}, laser diodes~\cite{cheng2014,vallet2016}, vertical external-cavity surface-emitting lasers (VECSEL) ~\cite{dumont2014}, fiber lasers~\cite{wangzheli2014,rota2014}, or dual-axis cavities~\cite{loas2014dual,danion2014}. A general issue concerning DFL-based opto-RF oscillators is the stabilization of their beat frequency, and it has been shown that an optical feedback loop containing a frequency-shifting element can be successfully used to this end~\cite{kervevan2007beat,thevenin2011beat}. In particular, the DFL beat-note can be phase-locked to a master RF oscillator, so that the frequency stability of the master is transferred to the optically carried signal. In this context, an interesting synchronization regime called \emph{bounded-phase} has been isolated~\cite{,kelleher2010phasor,thevenin2011resonance}, where the phase-locked state loses stability via a Hopf bifurcation. After the bifurcation, the system displays sustained amplitude and phase oscillations. The master and slave oscillators are not phase-locked anymore, yet they maintain frequency-locking~\cite{romanelli2014measuring,brunel2016}. Being linked to the existence of a Hopf bifurcation, this behavior is rather generic~\cite{kelleher2012bounded,li2013phase,barois2014frequency}. 
\par In the present work, we consider a DFL submitted to frequency-shifted feedback, i.e., an opto-RF oscillator, driven in a parameter region where it undergoes a subcritical bifurcation leading directly to a chaotic state, and investigate whether this situation is compatible with bounded-phase dynamics. Synchronization of chaotic lasers has already been studied, e.g. in spatially coupled~\cite{roy1994,deshazer2001} and injection locked~\cite{volodchenko2001phase} Nd:YAG, injection-locked Nd:YVO$_4$~\cite{uchida2000}, injection-locked diode~\cite{murakami2003,kim2006phase,aviad2008}, as well as gas~\cite{boccaletti2002experimental}, lasers. In those systems each laser's output was chaotic \emph{per se}, and the coupling had a regularizing effect, by inducing either phase or complete chaos synchronization. The case studied here is different in that our opto-RF oscillator is not, in itself, a chaotic oscillator. In the free-running regime, it has a regular behavior; it is the coupling that is responsible for the outbreak of chaos. Furthermore, in the cited previous studies, the authors were mostly interested in showing that synchronization occurred, 
without characterizing and discriminating phase-locking and bounded-phase regimes. In the present case, the heterodyne nature of our source readily offers experimental access to the phase difference, while in the previous works the phase had to be retrieved numerically from intensity time series~\cite{deshazer2001,volodchenko2001phase,kim2006phase}. By measuring phase noise spectra, in DFLs it is possible to obtain a precise experimental characterization and a quantitative comparison of the different synchronization regimes.
The aim of this letter is to show a regime characterized by chaotic oscillations not only of the amplitude, but also of the phase, and to demonstrate experimentally that frequency-locking can be maintained, i.e., that bounded-phase synchronization is possible also in the presence of chaos. Furthermore, we show that, in spite of the chaotic phase oscillations, the beat-note signal still keeps a good spectral purity at low offset frequencies. 
\begin{figure}[htbp]
	\centering
	\includegraphics{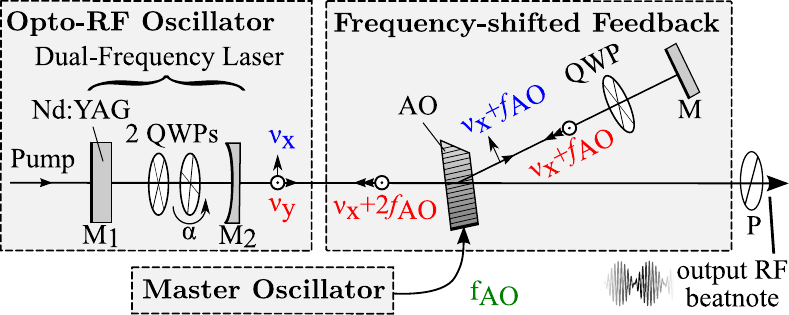}
	\caption{Experimental setup. The two polarization modes of a dual-frequency laser produce an optically-carried RF beat-note, which is locked to a master oscillator using frequency-shifted optical feedback. See text for details.} 
	\label{fig:setup}
\end{figure}
\par The experimental setup is similar to the one used in ~\cite{romanelli2014measuring} and is described in Fig.~\ref{fig:setup}. 
We study a solid-state Nd\textsuperscript{3+}:YAG laser oscillating simultaneously on two orthogonally-polarized modes. An intra-cavity tunable birefringence, realized using two quarter-wave plates (QWP), generates a frequency difference $\delta\nu = \nu_y-\nu_x$.
The interference of the two polarization modes after a polarizer $P$ produces an optically-carried RF beat-note, which we choose to tune around $180~\mathrm{MHz}$. We then add an external coupling between the two modes, through a reinjection of the field $E_x$ on the other one $E_y$, achieved using an external arm containing a QWP. Furthermore, the reinjected field is frequency-shifted using an acousto-optic modulator AO driven at the frequency $f_{AO}\approx 90~\mathrm{MHz}$ by a master RF oscillator. Provided that $\nu_x+2f_{AO}\approx \nu_y$, injection from $E_y$ into $E_x$ can be neglected because the frequency difference between $\nu_x$ and $\nu_y+2f_{AO}$ is much larger than the cavity bandwith. The optical coupling between the two modes is characterized by its strength (related to diffraction efficiency of the modulator, spatial mode-matching of the external cavity and other experimental factors) and by the detuning $\delta\nu-2f_{AO}$ between the RF source and the laser beat-note. The laser exhibits relaxation oscillations at a frequency $f_R=65~\mathrm{kHz}$, which allows to ignore delay effects in all optical paths, as this time scale is much longer than the round-trip time in the feedback arm, whose length is $70~\mathrm{cm}$.



\begin{figure}[tbp]
	\centering
	\includegraphics[width=7.6cm]{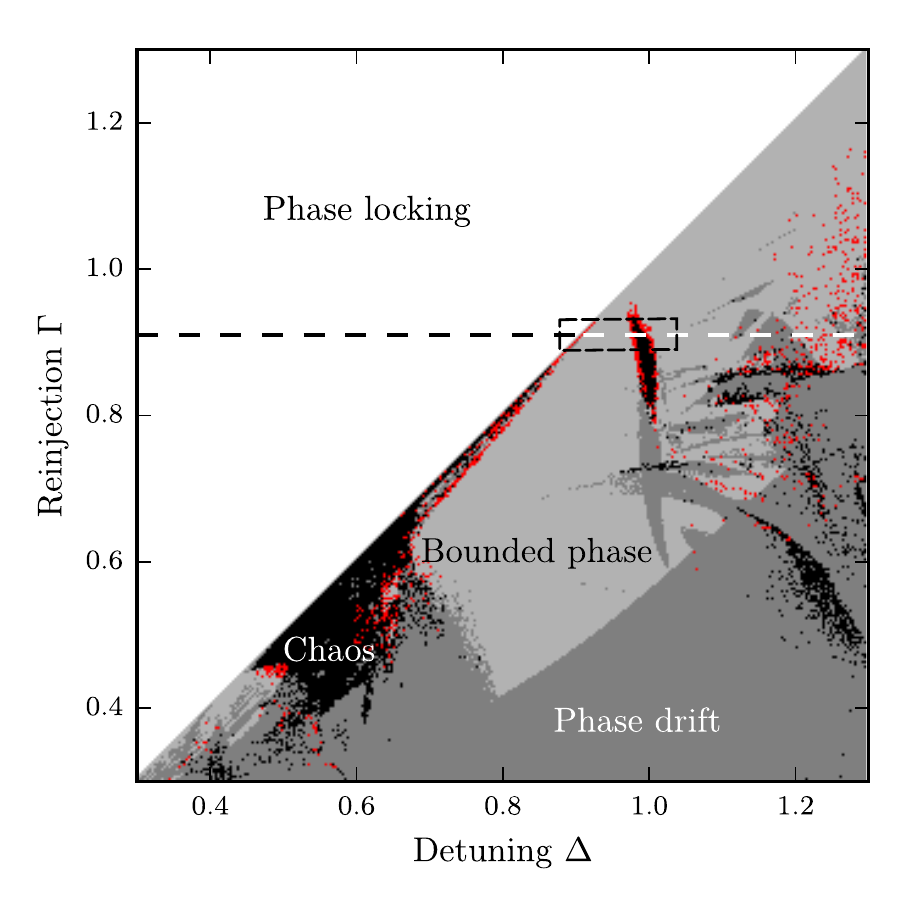}
	\caption{Map of different regimes of the system. Light gray: bounded phase (existence of at least a stable limit cycle for which $\phi$ is bounded), black: existence of at least one unbounded chaotic attractor, red: existence of bounded chaos. Dashed line $\Gamma=0.91$ is the line along which the bifurcation diagram of Fig.~\ref{fig:bifurc} is computed.} 
	\label{fig:map}
\end{figure}

\par We describe the system using the following rate equations for the electric fields and populations~\cite{thevenin_phase_2012}:
\begin{subequations}
	\label{eq:model}
	\begin{align}
		\frac{de_x}{ds}  =& \,\frac{m_x+\beta m_y}{1+\beta}\frac{e_x}{2} \\
		\frac{de_y}{ds}  =& \,\frac{m_y+\beta m_x}{1+\beta}\frac{e_y}{2} + i\Delta e_y + \Gamma e_x \\
		\begin{split}
			\frac{dm_{x,y}}{ds} =& \, 1- (|e_{x,y}|^2+\beta |e_{y,x}|^2) \\ &- \varepsilon m_{x,y} [1+(\eta-1)(|e_{x,y}|^2+\beta |e_{y,x}|^2)] 
		\end{split}
	\end{align}
\end{subequations}

\begin{figure}[tbp]
	\centering
	\includegraphics[width = 7.5cm]{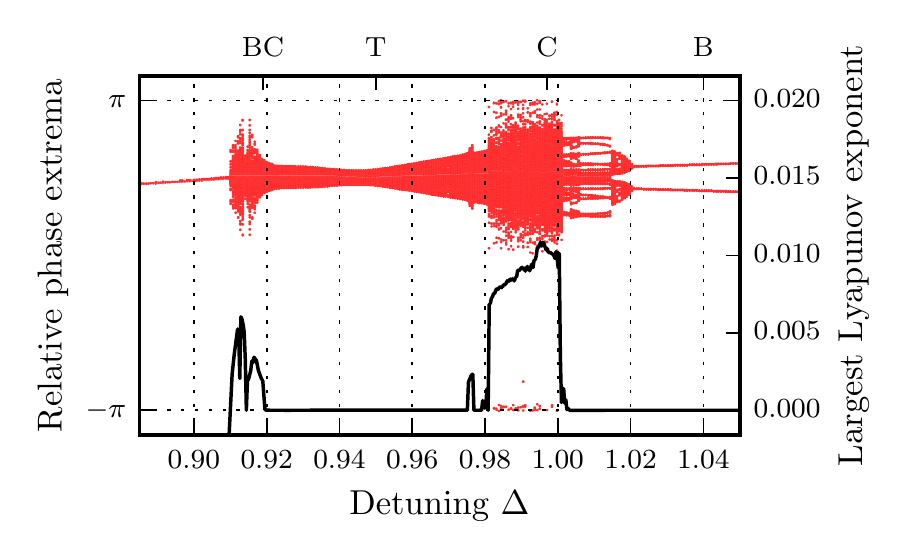}
	\includegraphics[width = 7.5cm]{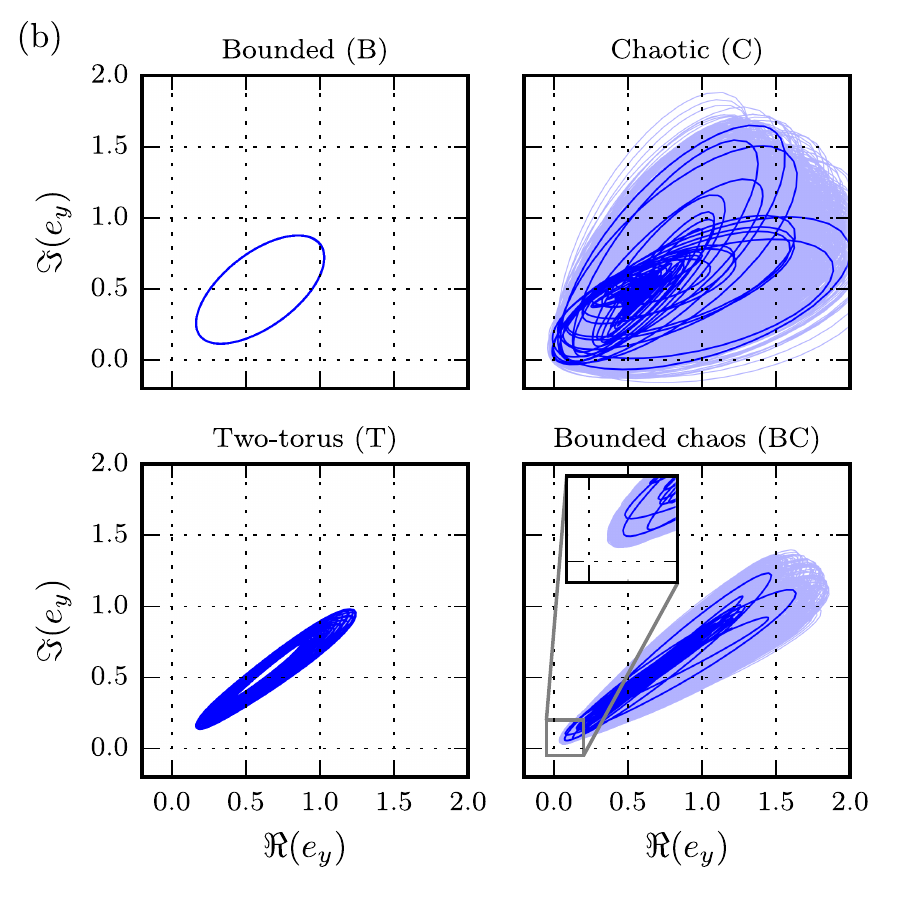}
	\caption{(a) Simulated bifurcation diagram of phase extrema for $\Gamma=0.91$. Final state at each point is taken as the initial conditions for the next point, as $\Delta$ is increased. The largest Lyapunov exponent is computed for each point and plotted as the solid black line. (b) Projection of the phase space trajectory on the $\{\Re(e_y),\Im(e_y)\}=\{|e_y|\cos\phi,|e_y|\sin\phi\}$ plane  for different regimes encountered (dark: 50000 time units, light: 2000).}
	\label{fig:bifurc}
\end{figure}

Here $e_{x,y}$ are the normalized amplitudes of the two complex fields $E_x = e_x \exp(2i\pi \nu_x t)$ and $E_y = e_y \exp(2i\pi(\nu_x+2 f_{AO})t)$. According to these equations, $e_x$ has a constant phase, and can thus be considered without loss of generality as a real quantity, so that only the phase $\phi$ of $e_y$ corresponds to the relative phase between the two oscillators. The normalized population inversions are $m_{x,y}$, and the normalized time $s = 2\pi f_R t$ is in units of the relaxation oscillations $f_R$. $\beta$ accounts for cross-saturation due to the spatial overlap in the gain medium, $\varepsilon$ for the gain medium losses, $\eta$ is the pump factor, $\Gamma$ quantifies the self-injection strength, and $\Delta=(\nu_y-\nu_x-2f_{AO})/f_R$ is the detuning between the laser beat-note and the RF driver. Without loss of generality, we consider $\Delta > 0$.
\par Eqs.~\ref{eq:model} are numerically integrated using an implicit Runge-Kutta method of order 5. Control parameters are $\Delta$ and $\Gamma$ while the others are fixed to $\beta=0.6$, $\eta=1.2$ and $\varepsilon=0.097$. These values have been previously measured~\cite{thevenin_phase_2012} and allow good model-experiment agreement. A mapping of the behavior of this dynamical system has been computed, unraveling chaotic, periodic, phase-bounded and phase-unbounded attractors, sometimes exhibiting multistability and fractal attraction basins.

Figure~\ref{fig:map} shows some of these features in a rich region of interest. Between the locking range $\Delta<\Gamma$, where the system reaches an equilibrium, and the unlocked region for $\Delta \gg \Gamma$, where the system settles on a limit cycle, various attractors exist and co-exist. It can be noted from this mapping that the domains of bounded phase behavior and of stability of any attractor other than fixed point are neither simple nor connected. It can also be noted that bounded-phase chaos, corresponding to the red points, is present only in small regions of the parameter space; nevertheless, as we will see, it can be clearly isolated experimentally.

\begin{figure}[htbp]
	\centering
	\includegraphics[width = 7.5cm]{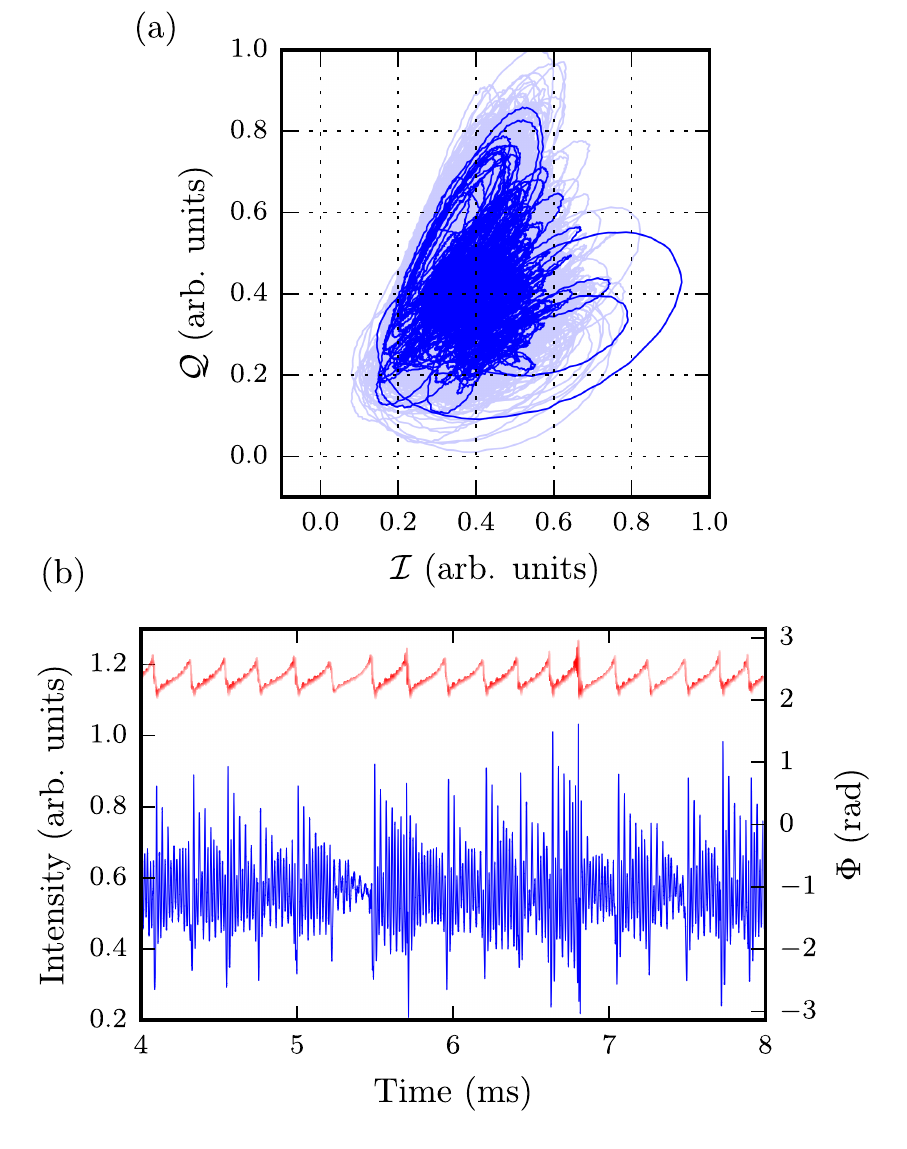}
	\caption{(a) Experimental phasor plot in the chaotic bounded phase regime (light: 100ms, dark: 10ms). (b) Corresponding time series of the intensity ($\sqrt{\mathcal{I}^2+\mathcal{Q}^2}$, blue lower trace) and phase ($\phi$, top red). $\mathcal{I}$ and $\mathcal{Q}$ are the two quadratures of the beat signal $I_{xy} = |E_x+E_y|^2 = |e_x|^2 + |e_y|^2 + 2|e_x e_y|\cos(2f_{AO}t+\phi)$, demodulated at the reference frequency $2 f_{AO}$, so that  $\mathcal{I}=2|e_x e_y|\cos\phi$ and $\mathcal{Q}=2|e_x e_y|\sin\phi$. The phase $\phi$ has the same meaning as the one in Fig~\ref{fig:bifurc}b.}
	\label{fig:boundedchaos}
\end{figure}

\par Figure~\ref{fig:bifurc} shows a higher definition slice of the map at $\Gamma=0.91$. On this phase bifurcation diagram, bounded phase regions are clearly seen, and different types of attractors are encountered as $\Delta$ changes. Starting from the right, for $\Delta > 1.02$, the system settles on a stable periodic orbit, which corresponds to the bounded phase regime (B). As $\Delta$ decreases and approaches the resonance with the relaxation oscillations, a Neimark-Sacker bifurcation leads to a torus, which is destabilized by other bifurcations in the multistable $0.97<\Delta<1.01$ region, where periodic and chaotic (C) attractors can be found. When $\Delta$ is further decreased away from resonance, the torus (T) is restored, before its chaotic break-up near the locking limit $\Delta=\Gamma$. In this particular region the phase remains bounded for a small range of detuning, where we witness a robust {\em bounded-phase chaos} regime (BC). Ultimately at the locking point, a subcritical Hopf bifurcation changes the stability of the fixed point, which becomes the only attractor in the phase space.

\par Knowing the parameters where it is supposed to arise, the previously unobserved bounded phase chaos could be investigated experimentally. Reinjection strength has been adjusted so that unlocking happens for $\delta\nu - 2 f_{AO}=0.9f_R$, thus ensuring $\Gamma=0.9$. Then detuning has been carefully set just after unlocking point, so that $\Delta \gtrapprox \Gamma$. Time series of length $100~\mathrm{ms}$ were recorded at a sampling rate of $10\,\mathrm{MHz}$ with a Rohde \& Schwarz FSV signal analyzer, performing IQ demodulation at $2f_{AO}$, thus allowing for a direct measurement of the quadratures of the beat-note signal, and of the relative phase. Results are shown in figure~\ref{fig:boundedchaos}, where we see a complicated trajectory in the IQ plane, corresponding to the projection of a strange attractor. The important point is that the trajectory never makes a loop around the origin, so that the phase variations, albeit chaotic, are relatively small.

Fig.~\ref{fig:boundedchaos} (b) presents a small portion of the measured time series, showing a spiking behavior, similar to the one expected from numerical simulations. The repetition rate looks quite regular, but the amplitude of the spikes is erratic. The phase, quickly increases during the spike then decreases before the next one, but as it remains bounded, the mean frequency of the opto-RF oscillator remains synchronized to the external RF source.

\begin{figure}[tbp]
	\centering
	\includegraphics[width = 8cm]{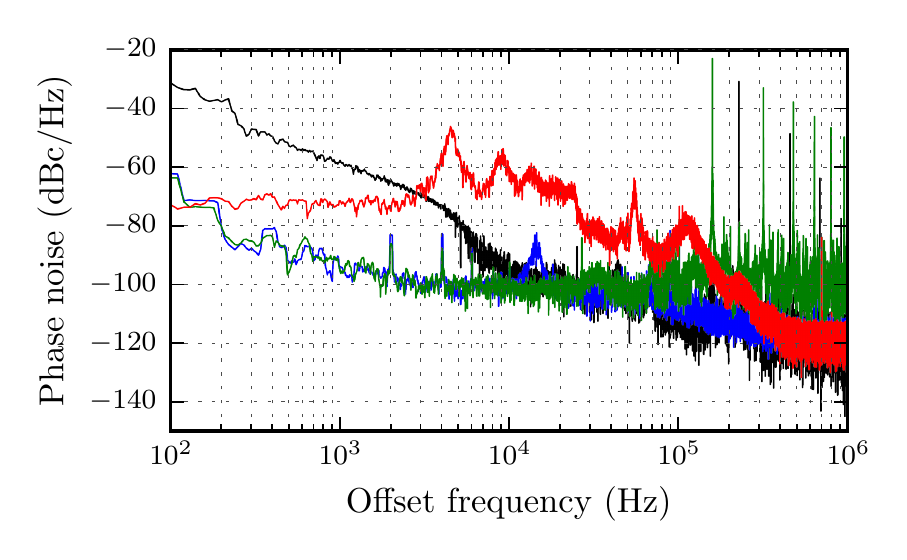}
	\caption{Experimental phase noise for different regimes at $\Gamma\approx 0.9$ (black: unlocked, blue: locked, green: bounded, red: bounded chaos).}
	\label{fig:phasenoise}
\end{figure}
\par But it is worth asking if we still have a good oscillator. For instance, it has been previously shown that even in the bounded phase regime, the quality and long-term stability of the master reference was transferred on the optical carrier~\cite{romanelli2014measuring}. This can be quantified by computing the power spectral density (PSD) of the relative phase. Experimental phase spectra extracted from the quadrature time series are presented in Fig.~\ref{fig:phasenoise}. It can be seen that the onset of chaos results in a broadening of the spectral peak around $50~\mathrm{kHz}$, corresponding to the value of $\Delta$, and of its harmonic at $100~\mathrm{kHz}$. 
\begin{figure}[htbp]
	\centering
	\includegraphics[width = 8cm]{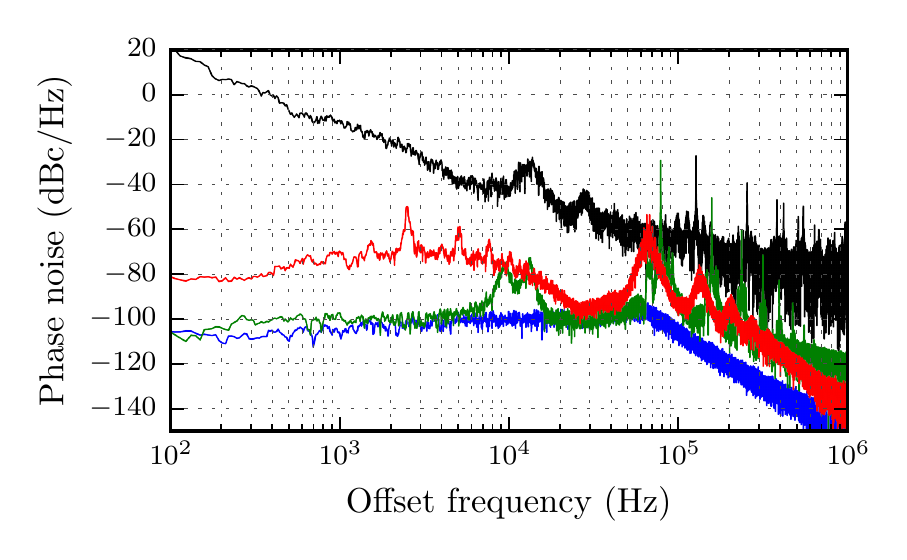}
	\caption{Simulated phase noise spectrum for different regimes at $\Gamma=0.9$ (black: drifting $\Delta=1.8$, blue: locked $\Delta=0.8$, green: bounded $\Delta=1.1$, red: bounded chaos $\Delta=0.91$). Computed by replacing $\eta$ with $\eta(1+0.03\xi(s))$, $\Delta$ by $\Delta(1+0.03\xi(s))$ and $\Gamma$ by $\Gamma(1+0.03\xi(s))$ where $\xi(s)$ is the same stochastic Gaussian process.}
	\label{fig:phasenoise-simu}
\end{figure}
A broad peak at low frequency, around $5$ kHz, also appears. 
Comparing with the time series in Fig.~\ref{fig:boundedchaos}, it can be seen that this peak corresponds to the average repetition rate of the spikes. 
In this part of the phase spectrum, bounded chaos features a higher phase noise level than all other curves, but it still outperforms clearly the free running laser at low frequency. In particular, we see that the low frequency part of the spectrum is flat, leading to a phase noise at $100~\mathrm{Hz}$ from the carrier which is comparable to the level of the phase-locking regime, and at least $30~\mathrm{dB}$ lower than the unlocked regime. Comparing with the free-running case, it appears clearly that in bounded-phase chaos the long-term synchronization of the mean value of the phase is well preserved.
\par For a qualitative comparison, the phase noise spectrum was computed from the numerical model, by introducing arbitrary gaussian white noise in the parameters $\eta$, $\Delta$ and $\Gamma$. The results, shown in Fig.~\ref{fig:phasenoise-simu}, are in good qualitative agreement with experiments. At $100~\mathrm{Hz}$ from the carrier, the simulated phase noise is lower for the phase-locking than for bounded-phase chaos, contrary to the experiments. This suggests that we have reached the noise floor of the experimental detection system. In any case, the simulations confirm the main features of the phase noise spectrum in the bounded-phase chaotic regime: a broadening of the peak corresponding to the detuning  frequency, and of its harmonics;  the appearance of a large, continuous component in the region between around $1~\mathrm{kHz}$ and $20~\mathrm{kHz}$, corresponding to the chaotic oscillations; and, most importantly, a flat spectrum in the $100~\mathrm{Hz}-1~\mathrm{kHz}$ region, with a noise level that is quite close to the phase-locking level, and in any case much lower than for the unlocked case. In short, it is confirmed that, when bounded-phase chaos occurs, synchronization is well preserved. A more quantitative agreement with the experiment would require a precise characterization of the noise sources of the experimental setup, beyond the scope of this work.

\par In summary, we have presented experimental evidence of bounded-phase chaos in a driven opto-RF oscillator destabilized by optical feedback. By measuring the phase noise spectra, we have shown that, in spite of this, the long-term phase stability of the master oscillator is still well transferred to the slave opto-RF oscillator. These features are well reproduced using a rate equation model. As it is the case for the non-chaotic bounded-phase regime~\cite{romanelli2014measuring}, we can expect these features to be common to a large class of systems. 
In particular, the effects of the Henry factor and of the feedback delay on this regime deserve to be investigated in opto-RF oscillators using semiconductor lasers~\cite{wang2014photonic}, for which chaotic sensing has been widely demonstrated~\cite{sciamanna2015physics}. The so-called {\em chaotic lidar}~\cite{lin2004chaotic} takes advantage of broadband chaos to achieve position sensing with a high resolution. Bounded-phase chaos, by imposing a good phase coherence to the chaotic waveform, will allow to combine this precise position sensing with the high resolution velocimetry allowed by phase coherence~\cite{vallet2013lidar}, i.e. to realize a {\em chaotic lidar-radar}.

\bibliography{biblio}

\end{document}